\documentclass[aps,prl,twocolumn,groupedaddress]{revtex4}

\usepackage{graphicx} 
\usepackage{dcolumn}
\usepackage{bm}

\begin{document}

\title{Polaronic high-temperature superconductivity in optimally doped bismuthate Ba$_{0.63}$K$_{0.37}$BiO$_{3}$ } 
\author{N.  Derimow$^{1}$, J. Labry$^{1}$, A. Khodagulyan$^{1}$,  J. Wang$^{2}$, and Guo-meng Zhao$^{1,*}$}
\affiliation{$^{1}$Department of Physics and Astronomy, 
California State University, Los Angeles, CA 90032, USA~\\
$^{2}$Department of Physics, Faculty of Science, Ningbo
University, Ningbo, P. R. China}

\begin{abstract}

Magnetic measurements have been carried out in the superconducting and normal states of the optimally doped nonmagnetic bismuthate superconductor Ba$_{0.63}$K$_{0.37}$BiO$_{3}$. The magnetic data along with previous $\mu$SR, resistivity, and tunneling data consistently show that there is a large polaronic enhancement in the density of states and effective electron-phonon coupling constant. The first-principle calculation within the density-functional theory indicates a small electron-phonon coupling constant of about 0.3-0.4, which can only lead to about 1~K superconductivity within the conventional phonon-mediated mechanism. Remarkably, the polaronic effect increases the electron-phonon coupling constant to about 1.4, which is large enough to leads to 32~K superconductivity. The present work thus uncovers the mystery of  high-temperature superconductivity in bismuthate superconductors, which will also provide important insight into the pairing mechanism of other high-temperature superconductors.

\end{abstract}
\maketitle 

The role of electron-phonon coupling in the pairing mechanism of high-temperature superconductivity in copper-based superconductors remains controversial although there have been compelling experimental \cite{ZhaoYBCO,McQueeney,Ros,Mih,Ved} and theoretical \cite{Review,Bauer} evidences for strong electron-phonon coupling and  for the existence of polaronic  supercarriers.  Because copper- and iron-based superconductors are in the proximity of antiferromagnetic  instability, it has been generally believed that antiferromagnetic fluctuation plays an essential role in bringing about high-temperature superconductivity in these two systems.  In contrast, high-temperature superconductivity in Ba$_{1-x}$K$_x$BiO$_3$ (BKBO) and MgB$_{2}$ cannot arise from antiferromagnetic fluctuation because they are not magnetic. The first-principle calculation of the superconducting transition in MgB$_{2}$ within the density-functional theory (DFT) and the multi-band anisotropic Eliashberg formalism \cite{Choi} can quantitatively explain the observed transition temperature, the isotope effect, phonon energy, and all other important physical properties.  This implies that the first-principle calculation within the DFT should be able to accurately predict electron-phonon coupling constant at least in nonmagnetic materials. On the other hand, the electron-phonon coupling constant of optimally doped BKBO is predicted to be about 0.3-0.4 from the first-principle calculation \cite{Ham,Mere}. This calculated electron-phonon coupling constant can only lead to about 1~K superconductivity within the single-band  Eliashberg formalism. Therefore, the conventional phonon-mediated theory is difficult to explain 30~K high-temperature superconductivity in the nonmagnetic BKBO.  One possible alternative mechanism is that pairing is mainly mediated by high-energy charge 
excitations~\cite{Batlogg}. This mechanism requires weak coupling so that the reduced energy gap 2$\Delta (0)/k_{B}T_{c}$ is  close to the 
value (3.53) predicted from the weak-coupling 
Bardeen-Cooper-Schrieffer (BCS) theory.  The second possible mechanism 
is that the effective retarded electron-phonon coupling constant $\lambda_{eff}$ increases significantly due to lattice polaronic effects \cite{Alex96,Zhao2001}.  Since the polaronic bandwidth is reduced, the effective density of states and the effective retarded electron-phonon coupling constant  increase by the polaronic enhancement factor $f_{p}$. This picture is consistent with the other independent theoretical studies of this system \cite{Allen,Fran,Kot}, which show that strong coupling to the high-energy oxygen breathing mode can lead to formation of polarons and even bipolarons. Remarkably, the  optical conductivity data of BKBO in both insulating and superconducting phases can be quantitatively explained \cite{Kot}.

The polaronic model predicts a large reduction in the optical Drude weight \cite{Kot} and a large enhancement in the effective density of states at the Fermi level $N^{*}(0)$ (Refs.~\cite{Alex96,Zhao2001}).  The Stoner enhancement is negligibly small in BKBO due to a simple $s-p$ hybridized conduction band and a large bonding length,  so the spin susceptibility of the conduction electrons is proportional to $N^{*}(0)$, which is enhanced by a factor of $f_{p}$ compared with the bare density of states $N_{b}(0)$. In contrast, within the conventional phonon-mediated mechanism,  the spin susceptibility is proportional to the bare density of states predicted from the DFT. Therefore,  precise determination of spin susceptibility for this material can make a clear distinction between the conventional phonon-mediated mechanism and the unconventional polaronic model.  Here we report measurements of the upper critical field, magnetic penetration depth, and normal-state susceptibility of  Ba$_{0.63}$K$_{0.37}$BiO$_{3}$. The current experimental results along with the previous tunneling spectrum \cite{Sam} and muon-spin-relaxation ($\mu$SR) \cite{ZhaoPRB2007} data consistently demonstrate the existence of a significant polaronic enhancement in the effective density of states, which leads to a huge enhancement in $T_{c}$ from about 1~K to 30~K.

\begin{figure}[htb]
     \vspace{-0.2cm}
    \includegraphics[height=11cm]{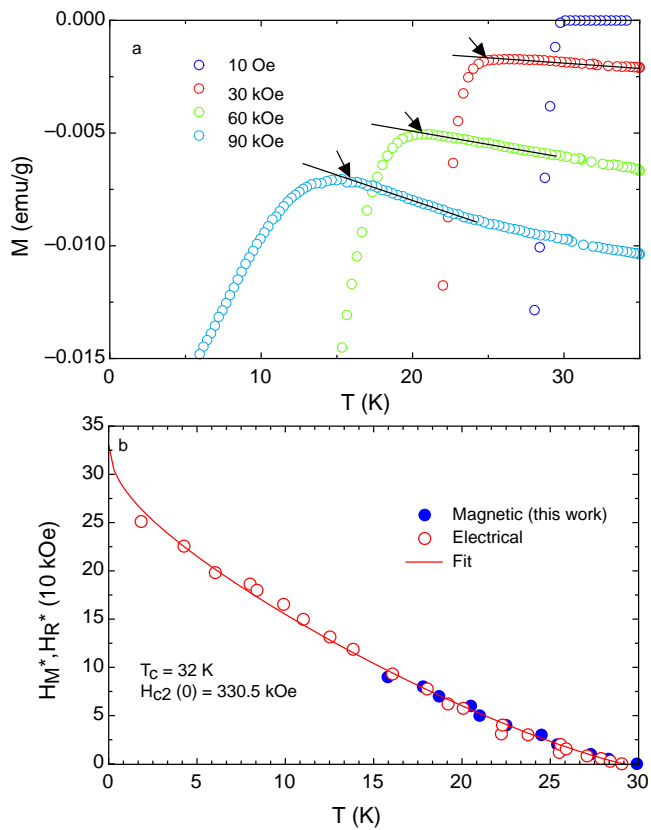}
     \vspace{-0.2cm}
 \caption[~]{a) Temperature 
dependencies of the field cooled magnetizations of the Ba$_{0.63}$K$_{0.37}$BiO$_{3}$ sample in different magnetic fields (up to 90 kOe). b) Temperature dependencies of critical fields measured magnetically $H_{M}^{*}$ and electrically $H_{R}^{*}$, respectively. The data for $H_{R}^{*}$ were taken from Ref.~\cite{Affronte}. The solid line is the curve with $T_{c}$ = 32~K and $H_{c2} (0)$ = 330.5$\pm$4.6 kOe, which was predicted from the model based on a large superconducting fluctuation \cite{Cooper}.  } 
\end{figure} 

Samples of Ba$_{0.63}$K$_{0.37}$BiO$_{3}$ were prepared by 
conventional solid state reaction following the same procedure as that reported in \cite{ZhaoPRB}. The detailed procedure is included in the Supplemental Material \cite{SM}.
Magnetizations were measured 
by a Quantum Design vibrating sample magnetometer. The absolute uncertainty of moment is less than 1$\times$10$^{-6}$ emu. Figure~1a shows the temperature 
dependencies of the field cooled magnetizations of the Ba$_{0.63}$K$_{0.37}$BiO$_{3}$ in different magnetic fields (up to 90 kOe). It is clear that the magnetic transition in the 10 Oe magnetic field is rather sharp. We define 
the critical temperature as the point of the onset of drop in magnetization (see arrows in Fig.~1a).
With this definition, we obtain phase diagram of the magnetically determined
critical field $H^{*}_{M}(T)$ for this bismuthate superconductor, as shown in Fig.~1b. It is interesting that the magnetically determined
critical field $H^{*}_{M}(T)$ coincides with the electrically determined
critical field $H^{*}_{R}(T)$ for a single-crystalline sample with a similar composition (Ref.~\cite{Affronte}).  This  
suggests that the critical fields obtained from both magnetic 
and electrical measurements are associated with the same 
physical phenomenon.  In contrast to the results for the conventional superconductors, the critical field curve of the bismuthate superconductor shows an upward curvature, which was well explained in terms of a large superconducting fluctuation proposed by Cooper {\em et al.} \cite{Cooper}. This model was also confirmed by the results reported for the sample irradiated by heavy ions \cite{Klein}. Because of the superconducting fluctuation, the critical fields determined from electrical and magnetic measurements are not true thermodynamic upper critical
fields at any finite temperature. Only the critical field at zero temperature is the true thermodynamic upper critical
field, which can be used to determine  the intrinsic thermodynamic quantities.   Because of the large superconducting fluctuation, the superconducting phase transition is not of second order, but of third or even fourth order \cite{Kumar,Hall}. The higher-order superconducting transition could make the specific-heat anomaly negligibly small, in agreement with experiments \cite{Gra,Wood}.

The solid line in Figure~1b is the best fitted curve by the equation derived from the theory of thermodynamic fluctuations \cite{Cooper}.  The zero-field $T_{c}$ is fixed to be 32~K for fitting and the best fit leads to the intrinsic zero-temperature upper critical field $H_{c2} (0)$ = 330.5$\pm$4.6 kOe.  At any finite temperature, the measured critical field is significantly suppressed compared with the intrinsic upper critical field.

\begin{figure}[htb]
     \vspace{-0.2cm}
    \includegraphics[height=11cm]{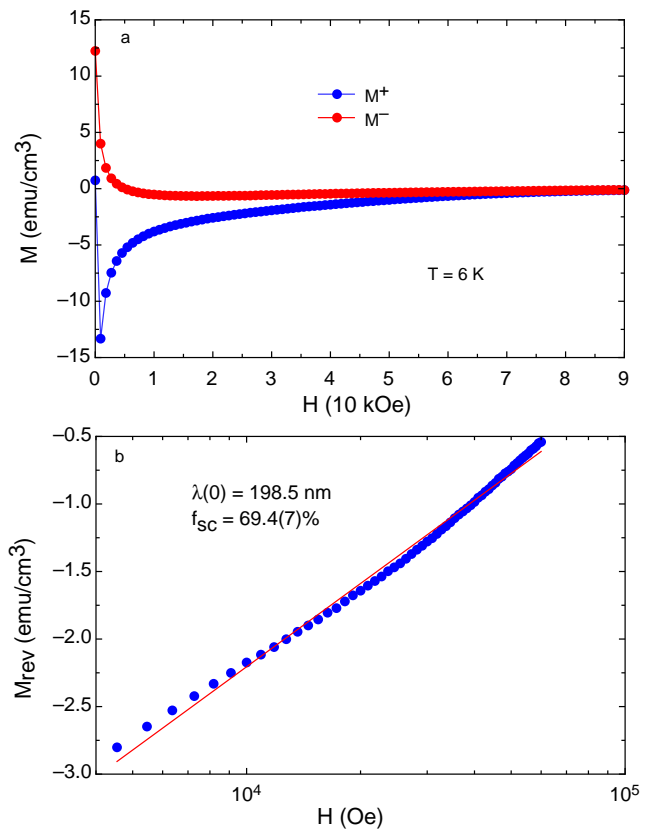}
     \vspace{-0.2cm}
 \caption[~]{a) Field dependence of the magnetization of the Ba$_{0.63}$K$_{0.37}$BiO$_{3}$ sample at 6~K. b) Field dependence of the reversible magnetization for the the sample at 6~K. The solid line is fitted by Eq.~\ref{M2} with a fixed magnetic penetration depth $\lambda (0)$ = 198.5~nm determined previously by $\mu$SR \cite{ZhaoPRB2007} and a fitting parameter of $f_{s}$ = 69.4$\pm$0.7$\%$.  } 
\end{figure} 

Figure 2a shows field dependence of the magnetization at 6.0 K.  The field-up magnetization $M^{+}$ is lower than the field-down magnetization $M^{-}$. The reversible magnetization $M_{rev} (H)$ can be calculated using the relation: $M_{rev} (H) = [M^{+}(H) + M^{-}(H)]/2$. According to the London model, the field dependence of the reversible magnetization is associated with the magnetic penetration depth $\lambda (T)$ as:
\begin{equation}\label{M1}
\frac{dM_{rev}(T, H)}{d\ln H}=\frac{f_{sc}\phi_{0}}{32\pi^{2}\lambda^{2} (T)},
\end{equation}
where $\phi_{0}$ is the quantum flux and $f_{sc}$ is the superconducting fraction. However, it was shown that the London model is quantitatively incorrect. 
More rigorous numerical calculation showed that in the case of the Ginzburg-Landau parameter $\kappa$ = 100 and in the field region of 0.02$\leq$ $H/H_{c2}$ $\leq$ 0.30, the London equation is modified as \cite{Hao}:
\begin{equation}\label{M2}
\frac{dM_{rev}(T, H)}{d\ln H}=\frac{0.77f_{sc}\phi_{0}}{32\pi^{2}\lambda^{2} (T)}.
\end{equation}
It is clear that the London model overestimates $\lambda (T)$ by a factor of 1.14.

In Figure 2b, we plot $M_{rev}$ versus magnetic field $H$ (in a logarithmic scale) in the field region of 0.02$\leq$ $H/H_{c2}$ $\leq$ 0.30. The solid line is the best linear fit to the data in this field regime. From the slope of the line and using  Eq.~\ref{M2} and $\lambda (6K)$ = $\lambda (0)$ = 198.5 nm (Ref.~\cite{ZhaoPRB2007}), we obtain $f_{sc}$ = 69.4$\%$, which is very close to that ($\sim$70$\%$) found from the $\mu$SR data \cite{ZhaoPRB2007}. The quantitative agreement between the $\mu$SR and magnetization data suggests that the obtained $\lambda (0)$ is reliable. 

\begin{figure}[htb]
     \vspace{-0.2cm}
    \includegraphics[height=5.5cm]{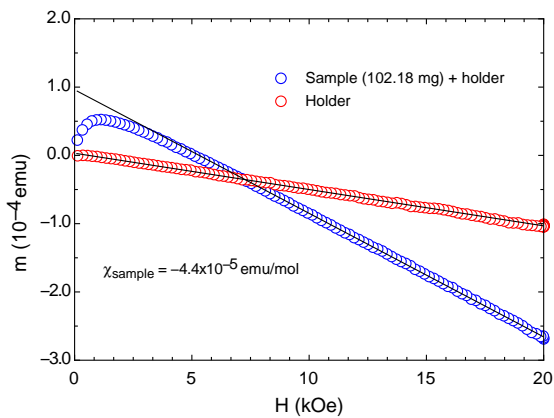}
     \vspace{-0.2cm}
 \caption[~]{Field dependencies of magnetic moments for the sample-holder alone and for the combined sample+sample-holder.  The solid lines are the best linear fits to the data above 5000~Oe.  The magnetic susceptibility of the sample alone is $-$4.4$\times$10$^{-5}$ emu/mol.    } 
\end{figure} 

In order to accurately determine the intrinsic normal-state susceptibility of the sample, we measure separately at 300~K the field dependencies of the moments for the sample-holder alone and for the combined sample+sample-holder. The data are shown in Fig.~3. Since the moment of any ferromagnetic/ferrimagnetic  impurity is saturated above 5000 Oe, linear fits to the $M$-$H$ curves above 5000 Oe yield the intrinsic $dm/dH$ values for the sample-holder alone and for the combined sample+sample-holder, respectively. The susceptibility of the sample is the difference in the $dm/dH$ values divided by the sample mass. Using this method, the intrinsic susceptibility is not influenced by possible presence of any ferromagnetic/ferrimagnetic impurity.  We have done several repeated measurements and found the same susceptibility value ($-$4.4$\times$10$^{-5}$ emu/mol) within 5$\%$.  This value lies in between two values obtained for two single-crystalline samples ($-$4.0$\times$10$^{-5}$ emu/mol for Ba$_{0.63}$K$_{0.37}$BiO$_{3}$ \cite{Bar} and $-$5.3$\times$10$^{-5}$ emu/mol for Ba$_{0.6}$K$_{0.4}$BiO$_{3}$ \cite{Hall}).  It is important to note that the subtraction of the sample-holder contribution is reliable only if the center positions of the sample-holder and the sample are the same. We have carefully checked the relative positions from the well-defined signals of the sample-holder and the combined sample+sample-holder. Therefore, our current measurements of the normal-state susceptibility should be reliable and accurate. 

With the reliably determined $\lambda (0)$ = 198.5 nm and $H_{c2} (0)$ = 330.5 kOe, we should be able to determine the thermodynamic quantities of the superconductor if we can accurately determine the reduced energy gap 2$\Delta (0)/k_{B}T_{c}$. Fortunately, a high-quality tunneling spectrum \cite{Sam} was measured earlier for BKBO with $T_{c}$ = 32~K. The point-contact tunneling spectrum is reproduced in Fig.~4. The solid line is the best fitted curve using the Blonder-Tinkham-Klapwijk (BTK) theory \cite{BTK}. The fitting parameters are displayed in the figure. From the inferred gap $\Delta (0)$ = 5.95 meV and the measured $T_{c}$ = 32~K, we obtain 2$\Delta (0)/k_{B}T_{c}$ = 4.31, indicating a quite large electron-boson coupling constant within the conventional strong-coupling theory. The result also  rules out the unconventional pairing mechanism based on the interaction with  high-energy charge excitations \cite{Batlogg}, which would predict 2$\Delta (0)/k_{B}T_{c}$ $\simeq$ 3.53. 

\begin{figure}[htb]
     \vspace{-0.2cm}
    \includegraphics[height=5.5cm]{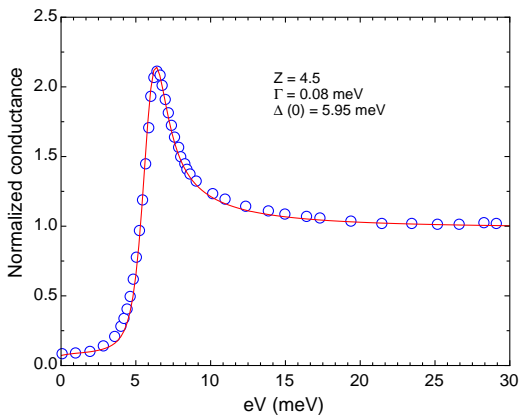}
     \vspace{-0.2cm}
 \caption[~]{Point-contact tunneling spectrum of optimally doped Ba$_{1-x}$K$_{x}$BiO$_{3}$ with $T_{c}$ = 32~K. The data were digitized from Ref.~\cite{Sam}. The solid line is fitted by the BTK theory with the following  fitting parameters: barrier strength $Z$ = 4.5, life-time broadening parameter $\Gamma$ = 0.08 meV, and zero-temperature superconducting gap $\Delta (0)$ = 5.95 meV.    } 
\end{figure} 

 From the reduced energy gap, we determine $k_{B}T_{c}/\hbar\omega_{\ln}$ to be 0.107 using a standard expression for conventional superconductors \cite{CarbotteRev}:
\begin{equation}\label{gap}
\frac{2\Delta (0)}{k_{B}T_{c}} = 3.53[1 + 
12.5(\frac{k_{B}T_{c}}{\hbar\omega_{ln}})^{2}\ln(\frac{\hbar\omega_{ln}}{2k_{B}T_{c}})].
\end{equation}
From $k_{B}T_{c}/\hbar\omega_{\ln}$ = 0.107,  $T_{c}$ = 32~K, and assuming a typical Coulomb pseudo-potential $\mu^{*}$ = 0.1, we obtain $\hbar\omega_{\ln}$ = 25.78 meV and $\lambda_{eff}$ = 1.41. The experimentally inferred numbers based on the conventional strong-coupling theory would imply the conventional phonon-mediated pairing mechanism in Ba$_{0.63}$K$_{0.37}$BiO$_{3}$.

The first problem with this conventional phonon mediated mechanism is that the inferred effective electron-phonon coupling constant is too large compared with the first-principle calculation based on the DFT. The second problem is that the electronic Sommerfeld coefficient $\gamma$ is calculated to be 8.18 mJ/molK$^{2}$ (see Supplemental Material \cite{SM}) from the unbiased parameters: $\lambda (0)$ = 198.5 nm, $H_{c2} (0)$ = 330.5 kOe, and  $k_{B}T_{c}/\hbar\omega_{\ln}$ = 0.107.  The inferred $\gamma$ value  is larger than the  bare Sommerfeld coefficient $\gamma_{b}$ (1.08 mJ/molK$^{2}$ for Ba$_{0.6}$K$_{0.4}$BiO$_{3}$ \cite{Mat1988}) by a factor of $f_{t}$ = 7.54. It is remarkable that this large enhancement factor $f_{t}$ = 7.54 is in quantitative agreement with the theoretically predicted  enhancement factor within the polaronic model \cite{Kot} (also see Supplemental Material \cite{SM}). In contrast,  this enhancement factor is  too large to be consistent with the conventional phonon-mediated mechanism. According to  the conventional model,  $\gamma = (1+ \lambda_{eff})\gamma_{b}$. This would imply $\lambda_{eff}$ = 6.54, in serous contraction with the inferred $\lambda_{eff}$ = 1.41 from the tunneling spectrum. Therefore, the conventional phonon mediated pairing mechanism cannot consistently explain the experimental results.

We can quantitatively explain the data in terms of a modified strong-coupling 
phonon-mediated mechanism \cite{Alex96,Zhao2001} where the strong electron-phonon coupling with
high-energy optical phonon modes leads to the formation of lattice polarons, and the 
polarons are bound into the Cooper pairs through the retarded 
electron-phonon interaction with other phonon modes.  Within 
this model the polaronic effect simply enhances the effective density of states by a 
factor of $f_{p}$  so that  the effective retarded 
electron-phonon coupling constant $\lambda_{eff}$ increases by the 
same factor, that is, $N^{*}(0)$ = $f_{p}N_{b}(0)$, $\lambda_{eff} = f_{p}\lambda_{b}$, where  $\lambda_{b}$ is the bare electron-phonon coupling constant determined from the first-principle calculation within the DFT. The total mass 
enhancement factor $f_{t}$ within this model is given by
\begin{equation}\label{FT}
f_{t} = f_{p} (1+\lambda_{eff}) = f_{p} (1+f_{p}\lambda_{b}). 
\end{equation}
The enhancement factor $1+\lambda_{eff}$ arises from the retarded 
electron-phonon interaction, which is treated within the Migdal approximation.

Substituting the value of $f_{t}$  = 7.54 and $\lambda_{eff}$ = 1.41 into Eq.~\ref{FT} yields $f_{p}$ = 3.26 and $\lambda_{b}$ = 0.40. The inferred $\lambda_{b}$ = 0.40 is in quantitative agreement with the first-principle calculation \cite{Ham}. 

The polaron mass enhancement factor $f_{p}$ of about 3.26 should be independently seen in the spin susceptibility $\chi_{s}$ in the normal state. Within the polaronic 
model,  $\chi_{s} = \mu_{B}^{2}N^{*}(0) =\mu_{B}^{2}N_{b}(0)f_{p}$. On the assumption that the nonsuperconducting phase is insulating and has zero spin susceptibility, the total susceptibility $\chi$ is given by
\begin{equation}
\chi = f_{sc}\chi_{s} (1-\frac{m^{2}_{e}}{3f^{2}_{p}m_{b}^{2}})+\chi_{core} + \chi_{para}(T),
\end{equation}
where $\chi_{core}$ is the core diamagnetic susceptibility and $\chi_{para}$ is the temperature dependent Curie-Weiss paramagnetic susceptibility. 
With $\chi_{core}$ = $-$7.8$\times$10$^{-5}$ emu/mol (Ref.~\cite{Batlogg}), $\chi_{para}(300K)$ = 0.33$\times$10$^{-5}$ emu/mol (which is estimated from the temperature dependent measurement and also very close to those for single-crystalline samples \cite{Hall}), $m_{b}$ = 0.647$m_{e}$ (Ref.~\cite{Mat1983}), and $f_{p}$ = 3.26, we obtain $\chi (300K)$ = $-$4.36$\times$10$^{-5}$ emu/mol, which is in excellent agreement with the measured value of $-$4.4$\times$10$^{-5}$ emu/mol. This quantitative agreement suggests that the inferred polaronic enhancement factor of 3.26 is reliable.

In summary, various experiments consistently demonstrate a significant enhancement in the density of states in nonmagnetic bismuthate superconductors due to the lattice polaronic effects.  The polaronic effect enhances the superconducting transition temperature from about 1~K  to 32~K. The present work thus uncovers the mystery of  high-temperature superconductivity in bismuthate superconductors, which will also provide important insight into the pairing mechanism of other high-temperature superconductors.

\noindent {\bf Acknowledgment:} 
ND acknowledges financial support from NIH and NIGMS under MBRS-RISE M.S.-to-Ph.D. Program (R25GM061331).

\noindent$^{*}$gzhao2@calstatela.edu


\begin{thebibliography}{99}

\bibliographystyle{prsty}
\bibitem{ZhaoYBCO}G. M. Zhao and D. E. Morris, Phys. Rev. B 
{\bf 51}, 16487(R) (1995); G. M. Zhao, M. B. Hunt, H. Keller, and K. A. 
M\"uller, Nature (London) {\bf 385}, 236 (1997); R. Khasanov, D. G. Eshchenko, H. Luetkens, E.
Morenzoni, T. Prokscha, A. Suter, N. Garifianov, M. Mali, J. Roos, K. Conder, and H. Keller, Phys. Rev. Lett. {\bf 92}, 
057602 (2004); A. S. Alexandrov and G. M. Zhao, New Journal of Physics {\bf 14}, 013046 (2012).
\bibitem{McQueeney}R. J. McQueeney, Y. Petrov, T. Egami, M. Yethiraj, 
G. Shirane, and Y. 
Endoh, Phys. Rev. Lett. \textbf{82},  628 (1999); D. Reznik, L. Pintschovius, M. Ito, S. Iikubo, M. Sato,
H. Goka, M. Fujita, K. Yamada, G. D. Gu, and J. M.
Tranquada, Nature (London) {\bf 440}, 1170 (2006).

\bibitem{Ros} O. R\"osch, O. Gunnarsson, X. J. Zhou, T. Yoshida, T.
Sasagawa, A. Fujimori, Z. Hussain, Z.-X. Shen, and S.
Uchida, Phys.  Rev.  Lett. \textbf{95} 227002 (2005); X. J. Zhou {\em et al.}, Phys.  Rev.  Lett.  {\bf 95}, 117001 (2005).

\bibitem{Mih} D. Mihailovic, C. M. Foster, K. Voss, and A. J. Heeger,
Phys. Rev. B 42, 7989 (1990); O. V. Misochko, E. Ya. Sherman, N. Umesaki, K. Sakai, S. 
Nakashima, Phys. Rev. B {\bf 59}, 11495 (1999); A. S. Mishchenko, N. Nagaosa, Z.-X. Shen, G.
De Filippis, V. Cataudella, T. P. Devereaux, C. Bernhard, K. W. Kim, and J. Zaanen,
Phys. Rev. Lett. {\bf 100}, 166401 (2008).

\bibitem{Ved} S. I. Vedeneev, P. Samuely, S. V. Meshkov, G. M.  
Eliashberg, A. G. M.  Jansen,  and P. Wyder, Physica C {\bf 198}, 47 (1992); D. Shimada, Y. Shiina, A. Mottate, Y. Ohyagi, and N.  
Tsuda, 
Phys. Rev. B {\bf 51}, 16495 (1995); R. S. Gonnelli, G. A. Ummarino, and V. A. Stepanov, Physica C {\bf 275}, 162 (1997); H. Shim, P. Chaudhari, G. Logvenov and I. Bozovic, Phys.
Rev. Lett. {\bf 101}, 247004 (2008); G. M. Zhao, Phys. Rev. Lett {\bf 103}, 236403 (2009).

\bibitem{Review}For a recent review see A. S. Alexandrov and J. T. Devreese,
{\em Advances in Polaron Physics} (Springer, Berlin
2009); A. S. Alexandrov and P. E. Kornilovitch, Phys. Rev. Lett. {\bf 82}, 807 (1999).
\bibitem{Bauer}T. Bauer
and C. Falter, Phys. Rev. B {\bf 80}, 094525 (2009).


\bibitem{Choi}H. J. Choi, D.  Roundy, H. Sun, M.  L. Cohen, and S.  G. Louie, Phys. Rev. B {\bf 66}, 020513(R) (2002).

\bibitem{Ham} N. Hamada, S. Massidda, and A. J. Freeman, Phys. Rev. B {\bf 40}, 4442 (1989). The electron-phonon coupling constant of Ba$_{0.7}$K$_{0.3}$BiO$_{3}$
is calculated to be 0.41 when one uses the Debye temperature $\theta_{D}$ = 513~K, which corresponds to $\hbar\sqrt{<\omega^{2}>}$ = 31.29 meV or $\hbar\omega_{\ln}$= 25.78 meV.
\bibitem{Mere}V. Meregalli and S.Y. Savrasov, Phys. Rev.
B {\bf 57}, 14453 (1998).

\bibitem{Batlogg} B.  Batlogg {\em et al.}, Phys.  Rev.  Lett.  {\bf 61}, 1670 
(1988).

\bibitem{Alex96} A. S. Alexandrov, V. V. Kabanov, Phys.  Rev.  B {\bf 54}, 
3655 (1996).

\bibitem{Zhao2001}G.  M.  Zhao, V.  Kirtikar, and D.  E.  Morris, Phys.  Rev.  B {\bf 63}, 220506(R) (2001).
\bibitem{Allen}I. B. Bischofs, V. N. Kostur, and P.  B. Allen, Phys. Rev. B {\bf 65}, 115112 (2002).

\bibitem{Fran}C. Franchini, G. Kresse, and R. Podloucky, Phys. Rev. Lett.  {\bf 102 }, 256402 (2009).


\bibitem{Kot} R. Nourafkan, F. Marsiglio, and G. Kotliar, Phys. Rev. Lett.  {\bf 109 }, 017001 (2012).

\bibitem{Sam}P. Samuely,  P. Szabo , A. G. M. Jansen, P. Wyder. J. Marcus,  C. Escribe-Filippini, and
M.Affronte, Physica B {\bf 194-196}, 1747 (1994).

\bibitem{ZhaoPRB2007} G. M. Zhao, Phys. Rev. B {\bf 76}, 020501(R) (2007).

\bibitem{ZhaoPRB}G.  M.  Zhao and D.  E.  Morris, Phys.  
Rev.  B \textbf{51}, 12 848 (1995). 

\bibitem{SM} Supplemental Material can be found at

\bibitem{Affronte} M.  Affronte {\em et al.}, Phys.  Rev.  B {\bf 
49}, 3502 (1994).

\bibitem{Cooper}J. R. Cooper, J. W. Loram, and J. M. Wade, Phys. Rev. B
{\bf 51}, 6179 (1995).

\bibitem{Klein} T. Klein, C. Marcenat, S. Blanchard,  J. Marcus, C. Bourbonnais, R. Brusetti,
C. J. van der Beek, and M. Konczykowski, Phys. Rev. Lett. {\bf 92}, 037005 (2004).


\bibitem{Kumar} P.  Kumar, D. Hall, and R. G. Goodrich, Phys. Rev. Lett., {\bf 82}, 4532 (1999).

\bibitem{Hall}D. Hall, R. G. Goodrich, C. G. Grenier, P.  Kumar, M.  Chaparala, and M. L. Norton, arXiv: cond-mat/9912160.

\bibitem{Gra}J. E. Graebner, L. F. Schneemeyer, and J. K. Thomas, Phys. Rev. B {\bf 39}, 9682 (1989).

\bibitem{Wood}B. F. Woodfield, D. A. Wright, R. A. Fisher, N. E. Philips, and H.Y. Tang, Phys. Rev. Lett. {\bf 83}, 4622 (1999).

\bibitem{Hao}Z. D. Hao and J. R. Chem, Phys. Rev. Lett. {\bf 67}, 2371 (1991).

\bibitem{Bar}S. N. Barilo, S. V. Shiryaev, and V. I. Gatalskaya, J. W. Lynn, M. Baran, H. Szymczak,  R. Szymczak, and D. Dew-Hughes, Phys. Rev. B {\bf 58}, 12355 (1998).

\bibitem{BTK} G. E. Blonder, M. Tinkham, and T. M. Klapwijk, Phys. Rev. B {\bf 25}, 4515 (1982).

\bibitem{CarbotteRev}J.  P.  Carbotte, Rev.  Mod.  Phys.  
{\bf 62}, 1027 (1990).  



\bibitem{Mat1988}L. F. Mattheiss and D. R. Hamann, Phys.  Rev.  Lett.  {\bf 60}, 2681 (1988).

\bibitem{Mat1983}L. F. Mattheiss, Phys. 
Rev. B \textbf{28}, 6629 (1983).












\end{thebibliography}
\end{document}